\documentclass[amsmath,amssymb,
showpacs,twocolumn, nobibnotes,aps,prb]{revtex4-1}

\usepackage{graphicx,amsmath,amssymb}
\usepackage{mathrsfs,eufrak}
\usepackage{dcolumn,color}
\usepackage{bm}
\usepackage{bbold}

\newcommand{\be}{\begin{equation}}
\newcommand{\ee}{\end{equation}}
\def\ba{\begin{aligned}}
\def\ea{\end{aligned}}

\newcommand{\bea}{\begin{eqnarray}}
\newcommand{\eea}{\end{eqnarray}}

\newcommand{\tr}{{\rm Tr}}

\begin{document}

\title{Multibody expansion of the local integrals of motion:
How many pairs of particle-hole do we really need to describe the quasiparticles in the many-body localized phase?} 
\author{Zahra  Gholami}
\author{Mohsen Amini}
\email{msn.amini@sci.ui.ac.ir}
\author{Morteza Soltani}  
\email{Author to whom correspondence should be addressed to. Electronic mail: mo.soltani@sci.ui.ac.ir}
\author{Ebrahim Ghanbari-Adivi}

\affiliation{Department of Physics, Faculty of Physics, University
of Isfahan, Isfahan 81746-73441, Iran}

\begin{abstract}
The emergent integrability in a many-body localized (MBL) system can be well characterized by the existence of the 
complete set of local integrals of motion (LIOMs). Such exactly conserved and exponentially localized operators are often understood as quasiparticle operators which can be expanded in terms of single-particle operators dressed with different numbers of particle-hole pairs. 
 Here, we consider a one-dimensional  XXZ spin-$\frac12$ Heisenberg chain in the presence of a random field and try to quantify the corrections needed to be considered in the picture of quasiparticles associated with LIOMs due to the presence of particle-hole excitations.  
 To this end, we explicitly present the multibody expansion of LIOM creation operators of the system in the MBL regime.
 We analytically obtain the coefficients of this expansion and discuss the effect of higher-order corrections associated with different numbers of particle-hole excitations.  
Our analysis shows that depending on the localization length of the system, there exist a regime in which the contributions that come from higher-order terms can break down the effective one-particle description of the LIOMs and such quasiparticles become essentially many-body-like.
 \end{abstract}
\maketitle
\section{Introduction~\label{Sec01}}
One of the best examples of the breaking of ergodicity is the many-body localization(MBL) phenomenon.  
It is a dynamical phase of disordered interacting quantum systems which fail to thermalize and hence challenges the very foundations of quantum statistical physics~\cite{RMP, Mirlin2005, BAA}. 
It is known as the many-body version of the Anderson localization of non-interacting disorder systems~\cite{Anderson} which is first predicted theoretically~\cite{BAA, Oganesyan, Luitz} and then observed experimentally mainly in cold-atom systems\cite{Ex1, Ex2, Ex3, Ex4, Ex5}.  
MBL is an interesting phase of the system as it exhibits many different characteristic features like
the exponentially small transport coefficients, retaining the local memory 
of the initial state for long times, and logarithmic-in-time growth of entanglement which are of particular importance~\cite{Oganesyan, Znidaric, Bardarson,Gopalakrishnan, Syrbin20}.
Indeed, these unique properties of the MBL phase are due to the existence of (quasi-)local operators that are exact integrals of motion and are usually known as LIOMs
~\cite{Serbyn13, Huse14, Chandran15, ROS, Imbrie17}.
To date, several different methods have been used to construct a complete set of LIOMs,
for instance, the construction of LIOMs via labeling the eigenstates through their corresponding LIOM eigenvalues~\cite{Serbyn13, Huse14}, by time-averaging local operators~\cite{Chandran15, Geraedts, Singh}, and different methods based on the exact diagonalization techniques~\cite{Rademaker, Abanin2016,Goihl, Peng, Amini2022}.
  
However, despite the vast progress made in developing different methods for the construction of LIOMs and the knowledge afforded by such investigations, 
quantifying their rich behavior is limited by a lack of explicit analytic studies of their physical properties.
In this sense, an important issue is to further understand the picture of the quasiparticles associated with LIOMs~\cite{Bera} which was originally pointed out by Basko, Aleiner, and Altshuler\cite{BAA}.
In this picture, LIOMs are quasiparticle density operators $n_i=\bar{c}^\dagger_i\bar{c}_i$ where $\bar{c}^\dagger_i=c^\dagger_i + \sum_{klm} A_{klm} c^\dagger_k c^\dagger_l c_m + \dots$ is a dressed version of some local products of single-particle operators $c^\dagger_i$.
In the absence of interaction (Anderson model), we have only the first term of this expansion which means that 
all the eigenstates are Slater determinants of single-particle states. However, in the presence of interaction, one needs to consider higher-order corrections 
comes from considering different numbers of pairs of particle-hole excitation in the above expansion.  
A natural fundamental question is then whether this expansion is perturbative. In a recent publication~\cite{Chen}, some aspects of this question are studied and it is shown that even if the individual eigenstates of the MBL system can be well approximated by single Slater determinants, they can still show deviations from the uniquely-defined single-particle picture which is itself another urgent reason for needing a better understanding of the picture of quasiparticles. 

In this work, we try to explicitly perform the above-mentioned expansion and obtain their coefficients in terms of different number of pairs of quasiparticles. 
We consider a disordered spin-$\frac12$ XXZ model on a one-dimensional lattice and focus on the effect of higher-order terms of the LIOMs expansion. 
For this system LIOM operators are a dressed version of some local products of the original Pauli operators and our analysis reveals that how the corrections associated with different number of particle-hole excitations my change the convergency of a perturbative expansion. Our results show that the convergence of such perturbative expansion strongly depends on the characteristic localization radius of the system in the MBL phase.

The rest of the paper is organized as follows.
In Sec.~\ref{Sec.II} we will draw the picture of quasiparticles associated with LIOMs precisely. For this purpose, we introduce our model first. Then, we will introduce a detailed analysis of the approach we used to obtain the explicit form of the coefficients of the expansion. 
Sec.~\ref{Sec.III} is devoted to presenting the numerical computations of the coefficients of the expansion and discussing its convergency for different localization regimes based on the localization length of the system and finally, concluding remarks are given in Sec.~\ref{Sec.IV}  
\section{The quasiparticle associated with LIOMs}\label{Sec.II}

\subsection{Model Hamiltonian}
We consider  a spin-$1/2$ XXZ chain of length $L$ with open boundary conditions in a 
random magnetic field in the $z$-direction which can be written as:  
\be\label{E1}
     \mathcal{H} =  J\left[ \sum_{i=1}^{L-1}   ({\sigma}^{x}_{i}{\sigma}^{x}_{i+1}+{\sigma}^{y}_{i}{\sigma}^{y}_{i+1})+\Delta {\sigma}^{z}_{i}{\sigma}^{z}_{i+1}+
    \sum_{i=1}^L h_{i}{\sigma}^{z}_{i}\right]
\ee
where $\sigma_i^{x,y,z}$ denote the Pauli operators acting on spin $i$.
We fix the exchange interaction coupling at $J=1$ as our unit of energy and choose the random fields $h_i$  independently  
from a random uniform distribution in  $[-W, W]$.
 The parameter $\Delta$ in this model determines the anisotropy in the interaction which is set to be $J_z=0$ for a free system and, $J_z\neq 0$  for an interacting 
 system.  
When $J_z=1$, the model is known to show a phase transition at a critical disorder strength $W=W_c\sim7$ 
from an ergodic phase to an MBL phase~\cite{PT1,PT2}. 
For the rest of the paper,  we will focus on the fully MBL side of the transition, $W>W_c$, in which the system fails to thermalize due to the existence of LIOMs.

\subsection{Expansion of the LIOMs in terms of pairs of particle-hole excitations}
As it is well known, one of the most important features of the  MBL regime is the existence of LIOMs  which allows to describe the system with an effective phenomenological Hamiltonian as
\be\label{E2}
     \mathcal{H} = \sum\limits_i {{\varepsilon _i}{{\tau} _{i}^z} + \sum\limits_{ij} {{\varepsilon_{ij}}} } {{\tau} _{i}^z}{\tau _{j}^z} + \sum\limits_{ijk} {{\varepsilon_{ijk}}} {{\tau} _{i}^z}{{\tau} _{j}^z}{{\tau} _{k}^z} + ...,
\ee
in which the emergent localized pseudo-spin operators  ${\tau} _{i}^z$ are
related to physical spin operators by a local unitary transformation as
\be\label{E3}
\tau_i^z=U\sigma_i^z U^\dagger.
\ee
Coefficients $\varepsilon_{ij...}$ in Eq.(~\ref{E2}) with units of energy can be determined by choosing a particular set of LIOMs in Eq.(~\ref{E3}).
There are different schemes proposed~\cite{Chandran15, Serbyn13, Huse14, Geraedts, Goihl, Peng, Amini2022} to find such sets of LIOMs which, in principle,
should satisfy the following conditions:
(i)  ${\tau _{i}^z}$ should be (quasi-)localized and independent,  (ii) they obey the Pauli spin algebra commutator relations, and  by definition, (iii) they are conserved quantities ($\left[ {{\tau _{i}^z},\mathcal{H}} \right] = 0$), and should be identified in such a way that their interaction range decays rapidly.

Let us now use the above-mentioned property (ii), and write the operators ${\tau} _{i}^z$  in terms of the pseudo-spin raising and lowering operators $\tilde{\tau}_{i}^{\pm}$ as
\be\label{E4}
\tau_{i}^{z}= \tilde{\tau}_{i}^{+} \tilde{\tau}_{i}^{-} - \tilde{\tau}_{i}^{-} \tilde{\tau}_{i}^{+}.
\ee
Here, $\tilde{\tau}_{i}^{+}$ is a pseudo-spin raising operator and can be expanded 
in terms of $n$-pairs of fermion-like operators,  $\tilde{\tau}_{i}^{+ (n)}$,  which belong to a sector with  $n$-excited spins as
\be\label{E5}
\tilde{\tau}_{i}^{+} = \tilde{\tau}_{i}^{+ (1)} + \tilde{\tau}_{i}^{+ (2)}  + \cdots + \tilde{\tau}_{i}^{+ (\mathcal{N})} = \sum_{n=1}^\mathcal{N}  \tilde{\tau}_{i}^{+ (n)}
\ee
where each term shows the contribution of  its corresponding sector and can be constructed from the original physical spin operators using the Jordan–Wigner transformation.
Let us write out the first few terms explicitly. The first term is the fraction of the operator norm that comes from the single-particle sector (the sector with single excited spin) and can written in terms of raising operators $\tilde{\sigma}_{j}^{+}$ with coefficients $\alpha_{j}$ as
\be\label{E6}
\tilde{\tau}_{i}^{+ (1)} = \sum_{j} \alpha_{j}^{i} \tilde{\sigma}_{j}^{+},
\ee
in which  the $\tilde{\sigma}_{j}^{+}$ operator is related to the physical spin operators by the following Jordan–Wigner transformation
\be\label{E7}
\tilde{\sigma}_{j}^{+}=\left(\prod_{k<j} \sigma_{k}^{z}\right) \sigma_{j}^{+}.
\ee
Likewise, the second term in Eq.(~\ref{E5}) is the contribution of the two-particle sector to $\tilde{\tau}_{i}^{+}$ and reads as
\be\label{E8}
\tilde{\tau}_{i}^{+ (2)} = \sum_{j_{1}<j_{2}<j_{3}} \alpha_{j_{1},j_{2},j_{3}}^{i} \tilde{\sigma}_{j_{1}}^{+} \tilde{\sigma}_{j_{2}}^{+} \tilde{\sigma}_{j_{3}}^{-},
\ee
which represents the fraction of the operator norm that comes from two-particle sector of the Hilbert space correspondingly.

In general, the $k$-th term on the right-hand side of the Eq.(~\ref{E5}) can be treated as
\be\label{E9}
\tilde{\tau}_{i}^{+ (k)} = \sum_{j_{1},\cdots,j_{2k-1}} \alpha_{j_{1},\cdots,j_{2k-1}}^{i} \tilde{\sigma}_{j_{1}}^{+} \cdots \tilde{\sigma}_{j_{k}}^{+} \tilde{\sigma}_{j_{k+1}}^{-} \cdots \tilde{\sigma}_{j_{2k-1}}^{-}.
\ee 
In what  follows we will try to obtain the set of expansion coefficients $\alpha$ introduced in the above Eqs.~\eqref{E6} ,\eqref{E8}, and  \eqref{E9}. 

\subsection{Finding the coefficients of the expansion}
Before proceeding to obtain the coefficients of the expansion, it will be well to introduce the following notation in which the physical spin basis in the single-excitation sector is shown by
$|j\rangle\equiv|\sigma_j^+|0\rangle$ where the reference state $|0\rangle$ is considered as a state with all spins down and $j$ denotes the site with flipped spin.
By the same token, we can use $|j_1 j_2 \dots j_k\rangle\equiv|\sigma_{j_1}^+\sigma_{j_2}^+\dots\sigma_{j_k}^+|0\rangle$ to show a generic basis state in a sector with $k$-excitations. 
Similarly, we use the notation $\overline{|{j_1} {j_2} \dots {j_k}\rangle}\equiv|\tau_{j_1}^+\tau_{j_2}^+\dots\tau_{j_k}^+|0\rangle$ to show the same thing in the LIOM basis.

Let us suppose now that we have obtained the set of LIOM operators for this model. This can be done by employing different schemes introduced for this purpose~\cite{Goihl, Peng}, however, we will use the fast and accurate method in Ref.~\cite{Amini2022} which is recently discussed by our group. 
Now we can start the calculation of the coefficients sector by sector. 

In the single-excitation sector and as long as the LIOMs are available in this sector the following relation between the real space physical spin basis and LIOM basis set holds:
\be
\label{E10}
\overline{|\dot{\imath}\rangle} = \sum_{j} \beta_{j}^{i} |j\rangle,
\ee
in which $\beta_{j}^{i}$ are the matrix elements of the unitary  transformation $U=\sum_{j} \overline{|j\rangle} \langle j |$.
On the other hand, based on the above definition we have
\be
\label{E11}
\overline{|{\dot{\imath}}\rangle} = \tilde{\tau}_{i}^{+} |0\rangle=\tilde{\tau}_{i}^{+ (1)} |0\rangle = \sum_{j} \alpha_{j}^{i} |j\rangle
\ee
where we have used ~\eqref{E5} and ~\eqref{E6} and the fact that the terms with $n>1$ give zero when acting on the vacuum state $|0\rangle$, that is 
$\tilde{\tau}_{i}^{+ (n>1)} |0\rangle=0$. 
A comparison between Eqs.~\eqref{E10} and~\eqref{E10} immediately gives the coefficients $\alpha_{j}^{i}$ of the single-excitation sector as
\be
\label{E12}
\alpha_{j}^{i} = \beta_{j}^{i}.
\ee
Before going to the next sector it is worth mentioning that $\tilde{\tau}_{i}^{+ (1)}$ 
satisfies  the fermionic anticommutation relations
while the higher-order terms in  Eq.~\eqref{E5}  does not obey the fermionic algebra.

The next sector after the single-spin-flip sector has two flipped spins (two-excitation sector) for which the following relation holds
\be
\label{E13}
\overline{|{\dot{\imath}}_{1}\,{\dot{\imath}}_{2}\rangle} = \sum_{j_{2}>j_{1}} \beta_{j_{1},j_{2}}^{i_{1},i_{2}} |j_{1}\,j_{2}\rangle,
\ee
in which we need to keep  $ i_{2}>i_{1} $ and $ j_{2}>j_{1} $ to ensure the correct fermionic anticommutation relations.
Like the single-excitation sector, one can write the following relation using Eq.~\eqref{E8} 
\be
\label{E14}
\overline{|{\dot{\imath}}_{1}\,{\dot{\imath}}_{2}\rangle} = \tilde{\tau}_{i_{1}}^{+} \tilde{\tau}_{i_{2}}^{+} |0\rangle=(\tilde{\tau}_{i_{1}}^{+ (2)} + \tilde{\tau}_{i_{1}}^{+ (1)}) \tilde{\tau}_{i_{2}}^{+ (1)} |0\rangle
\ee
It is now possible to multiply the left side of the Eqs.~\eqref{E13} and ~\eqref{E14} by $\langle j_{01}\,j_{02}|$ which results in
\be
\label{E15}
\langle j_{01}\,j_{02}|(\sum_{j_{1},j_{2},j_{3}} \alpha_{j_{1},j_{2},j_{3}}^{i} \tilde{\sigma}_{j_{1}}^{+} \tilde{\sigma}_{j_{2}}^{+} \tilde{\sigma}_{j_{3}}^{-})\sum_{j^{\prime}} \alpha_{j^{\prime}}^{i} |j^{\prime}\rangle = \beta_{j_{01},j_{02}}^{\prime\,i_{1},i_{2}}
\ee
in which
\be
\label{E16}
\beta_{j_{01},j_{02}}^{\prime\,i_{1},i_{2}} = \beta_{j_{01},j_{02}}^{i_{1},i_{2}} - \langle j_{01}\,j_{02}| \tilde{\tau}_{i_{1}}^{+ (1)} \tilde{\tau}_{i_{2}}^{+ (1)} |0\rangle.
\ee
Thus one can easily conclude that:
\be
\label{E17}
\sum_{j^{\prime}} \alpha_{j_{01},j_{02},j^{\prime}}^{i_{1}} \beta_{j^{\prime}}^{i_{2}} = \beta_{j_{01},j_{02}}^{\prime\,i_{1},i_{2}}.
\ee
It is now possible to use the orthogonality condition
\be
\label{E18}
\sum_{i_{2}} \beta_{j^{\prime}}^{i_{2}} \beta_{j_{03}}^{\ast\,i_{2}} = \delta_{j^{\prime},j_{03}}
\ee
to obtain the coefficients $ \alpha_{j_{01},j_{02},j_{03}}^{i_{1}} $ as
\be
\label{E19}
\alpha_{j_{01},j_{02},j_{03}}^{i_{1}} = \sum_{i_{2}} \beta_{j_{01},j_{02}}^{\prime\,i_{1},i_{2}} \beta_{j_{03}}^{\ast\,i_{2}},
\ee
which shows that in order to obtain the coefficients of the expansion $\tilde{\tau}_{i}^{+ (2)}$  one needs only to use the information of the first and second sectors.

Similarly, for the three-particle sector, we have:
\be
\label{E20}
\overline{|{\dot{\imath}}_{1}\,{\dot{\imath}}_{2},{\dot{\imath}}_{3}\rangle} = \sum_{j_{3}>j_2>j_1} \beta_{j_{1},j_{2},j_{3}}^{i_{1},i_{2},i_{3}} |j_{1}\,j_{2}\,j_{3}\rangle,
\ee
and on the other hand,
\be
\label{E21}
\overline{|{\dot{\imath}}_{1}\,{\dot{\imath}}_{2},{\dot{\imath}}_{3}\rangle} = \tilde{\tau}_{i_{1}}^{+} \tilde{\tau}_{i_{2}}^{+} \tilde{\tau}_{i_{3}}^{+} |0\rangle = \tilde{\tau}_{i_{1}}^{+} (\sum_{j^{\prime}_{1},j^{\prime}_{2}} \beta_{j^{\prime}_{1},j^{\prime}_{2}}^{i_{2},i_{3}} |j^{\prime}_{1}\,j^{\prime}_{2}\rangle).
\ee
According to the expansion of $\tilde{\tau}_{i_{1}}^{+}$ , we conclude that
\be
\label{E22}
\langle j_{01}\,j_{02}\,j_{03}|\tilde{\tau}_{i_{1}}^{+ (3)}(\sum_{j^{\prime}_{1},j^{\prime}_{2}} \beta_{j^{\prime}_{1},j^{\prime}_{2}}^{i_{2},i_{3}} |j^{\prime}_{1}\,j^{\prime}_{2}\rangle) = \beta_{j_{01},j_{02},j_{03}}^{\prime\,i_{1},i_{2},i_{3}},
\ee
in which
\bea
\label{E23}
\beta_{j_{01},j_{02},j_{03}}^{\prime\,i_{1},i_{2},i_{3}} &=& \beta_{j_{01},j_{02},j_{03}}^{i_{1},i_{2},i_{3}}  \\
&-& \langle j_{01}\,j_{02}\,j_{03}| (\tilde{\tau}_{i_{1}}^{+ (1)} + \tilde{\tau}_{i_{1}}^{+ (2)})(\tilde{\tau}_{i_{2}}^{+ (1)} + \tilde{\tau}_{i_{2}}^{+ (2)})(\tilde{\tau}_{i_{3}}^{+ (1)})|0\rangle,\nonumber
\eea
and results in
\be
\label{E24}
\sum_{j_{1}^{\prime},j_{2}^{\prime}} \alpha_{j_{01},j_{02},j_{03},j_{1}^{\prime},j_{2}^{\prime}}^{i_{1}} \beta_{j_{1}^{\prime},j_{2}^{\prime}}^{i_{2},i_{3}} = \beta_{j_{01},j_{02},j_{03}}^{\prime\,i_{1},i_{2},i_{3}}.
\ee
Now it is again straightforward  to use the following orthogonality condition 
\be
\label{E25}
\sum_{i_{2},i_{3}} \beta_{j_{1}^{\prime},j_{2}^{\prime}}^{i_{2},i_{3}} \beta_{j_{04},j_{05}}^{\ast\,i_{2},i_{3}} = \delta_{j_{1}^{\prime},j_{04}} \delta_{j_{2}^{\prime},j_{05}}
\ee
to obtain the corresponding coefficients of this sector as
\be
\label{E26}
\alpha_{j_{01},j_{02},j_{03},j_{04},j_{05}}^{i_{1}} = - \sum_{i_{2},i_{3}} \beta_{j_{01},j_{02},j_{03}}^{\prime\,i_{1},i_{2},i_{3}} \beta_{j_{04},j_{05}}^{\ast\,i_{2},i_{3}}.
\ee
Again, it is obvious that the coefficients of the expansion, up to the third order, only depend on the information of the LIOMs in the sectors with the number of excitations equal to or smaller than this order. 

Finally, let us write the generic expression of the coefficients for an expansion up to an arbitrary order $L$ as
\be
\label{E27}
\alpha_{j_{1},\cdots,j_{2L-1}}^{i_{1}} = \sum_{i_{2},\cdots,i_{L}} \beta_{j_{1},\cdots,j_{L}}^{\prime\,i_{1},\cdots,i_{L}} \beta_{j_{L+1},\cdots,j_{2L-1}}^{i_{2},\cdots,i_{L}},
\ee
where:
\begin{widetext}
\begin{equation}
\label{E28}
\beta_{j_{1},\cdots,j_{L}}^{\prime\,i_{1},\cdots,i_{L}} = \beta_{j_{1},\cdots,j_{L}}^{i_{1},\cdots,i_{L}} - \langle j_{1},\cdots,j_{L}| (\sum_{m=1}^{L-1} \tilde{\tau}_{i_{1}}^{+ (m)})(\sum_{m=1}^{L-1} \tilde{\tau}_{i_{2}}^{+ (m)})(\sum_{m=1}^{L-2} \tilde{\tau}_{i_{3}}^{+ (m)})\cdots(\tilde{\tau}_{i_{L}}^{+ (1)})|0\rangle.
\end{equation}
\end{widetext}
In the next section, we will use the above-mentioned coefficients and confirm their validity by numerical computations.
 

\section{Numerical results}               \label{Sec.III}
 
To illustrate the accuracy of our theoretical analysis presented above we will now discuss some numerical 
results for a spin chain with different lengths $L=7,9,11$ and in the strong disorder regime.
Since we are interested in the full MBL regime we don't need to be much worry about the small system sizes and use the following strategy. 
We take sufficiently large disorder intensities, $W\gg W_c$, in such a way that the localization length $\xi$ is smaller than our system size, that is $\xi\ll L$ (to determine the localization length as a function of disorder intensity, $\xi(W)$, we can use the results of Refs.~\cite{Chandran15, Amini2022}). 
In what follows we consider large number of different realizations of disorder and take the LIOM located on the central site of 
the chain and will focus on two different quantities.
The first one is the $\tilde{\tau}_i^+$ defined in Eq.~\eqref{E5}.
We are interested to compare the resulting approximated-operator $\tilde{\tau}_i^+(\mathcal{N})$ truncated at different orders $\mathcal{N}$ with its exact numerical value $\tilde{\tau}_{i(\text{Exact})}^+$.  
Thus, we can use the method of Ref.~\cite{Amini2022} to obtain $\tilde{\tau}_{i(\text{Exact})}^+$ and then
define the relative norm difference between  this quantity and its approximated value  as
\be
\label{E29}
\mathcal{E}(\mathcal{N})=\frac{||\tilde{\tau}_i^+(\mathcal{N})-\tilde{\tau}_{i(\text{Exact})}^+||^2_F}{||\tilde{\tau}_{i(\text{Exact})}^+||^2_F},
\ee
where $||\mathcal{O}||^2_F=\tr(\mathcal{O}^\dagger \mathcal{O})$ defines the Frobenius norm of an operator $ \mathcal{O}$.

 \begin{figure}[t!]
	\center{\includegraphics[width=1.0\linewidth]{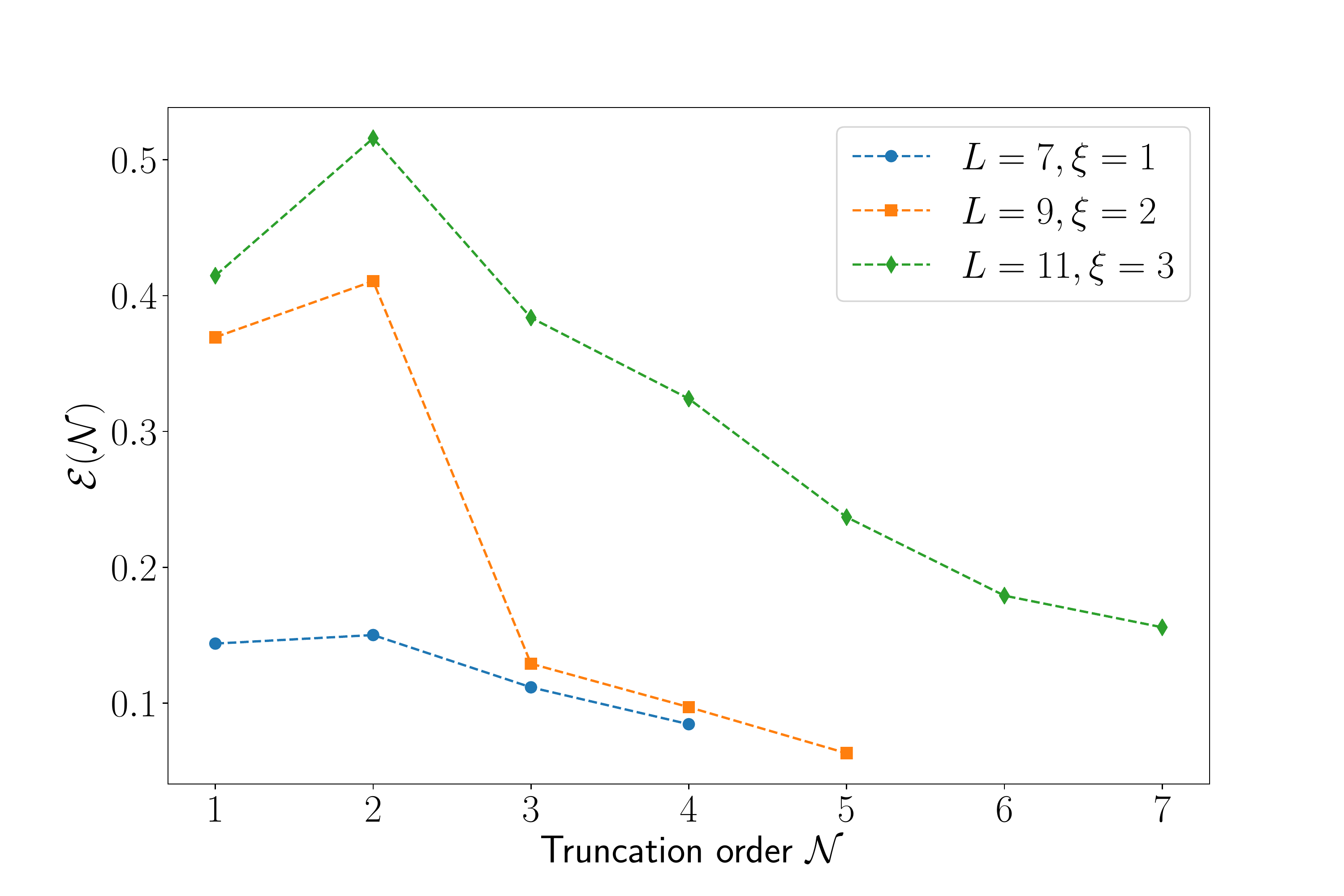}} 
	\caption{Difference between the corrections comes from different excitation sectors in the expansion of the LIOM creation operator $\tilde{\tau}_i^+(\mathcal{N})$
 truncated at different orders $\mathcal{N}$ and its exact numerical value which is defined as $\mathcal{E}(\mathcal{N})$ in Eq.~\eqref{E29} for a spin chain of lengths $L=7,9,11$ and with different localization lengths $\xi=1,2,3$  correspondingly which is averaged over different disorder configurations.}
	\label{FG1}
\end{figure}

Fig.~\ref{FG1} represents the relative norm difference $\mathcal{E}$ defined in Eq.~\eqref{E29} as a function of the truncation order of the expansion $\mathcal{N}$.
It is obvious that the effects of higher-order corrections strongly depend on the order of expansion. This means that they are not necessarily a monotonic decreasing function of the order of expansion. As we can see, when the order of truncation is smaller than the localization length $\xi$ measured in the lattice spacing units $a$, ($\mathcal{N}<(\xi/a)$),
the corrections associated with higher-order become important and not negligible. In contrast, when the truncation order reaches the localization length ($\mathcal{N}>(\xi/a)$), higher-order corrections become a monotonic decreasing function of the truncation order of the expansion. 

\begin{figure}[t!]
	\center{\includegraphics[width=1.0\linewidth]{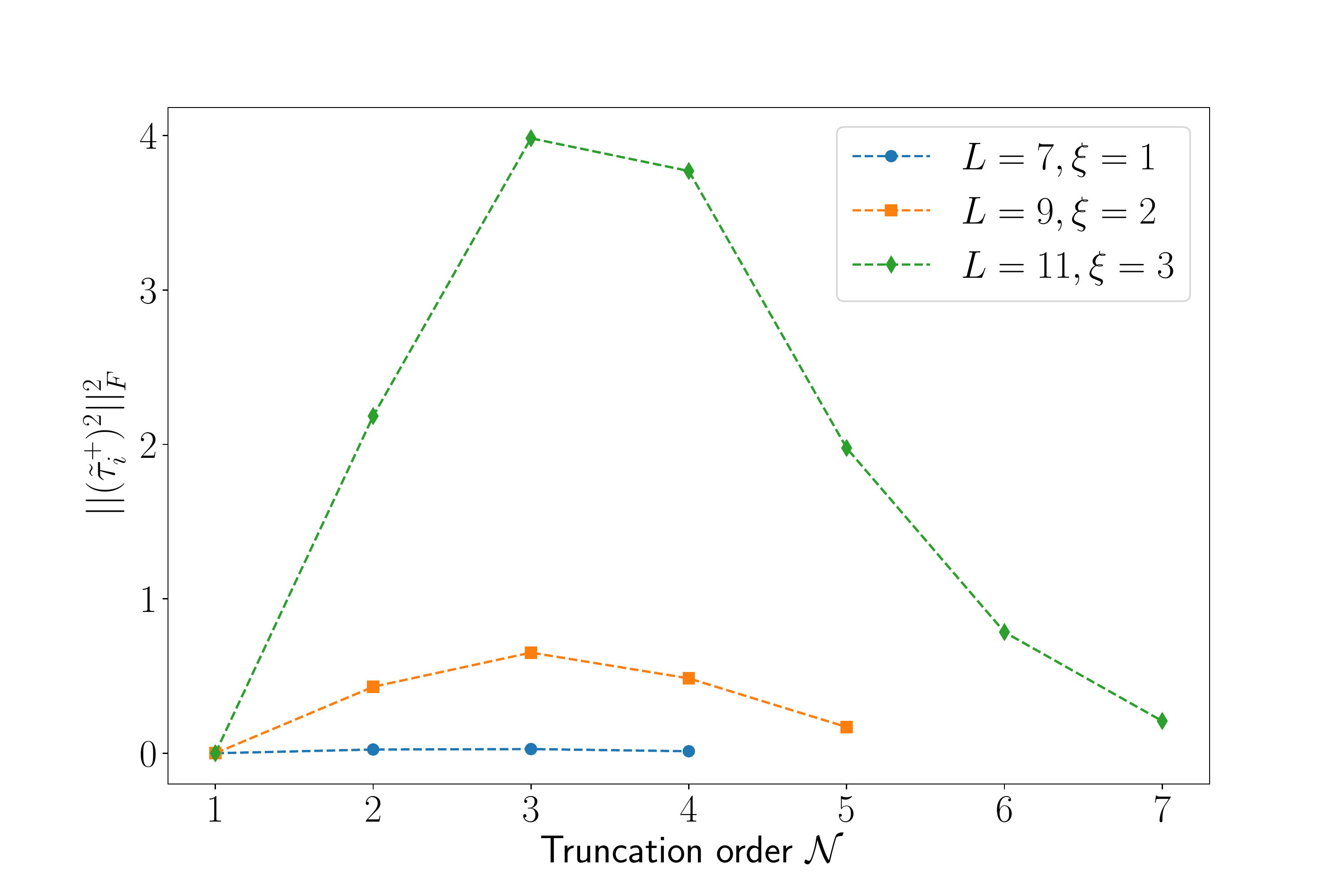}} 
	\caption{The Frobenius norm of the operator $(\tilde{\tau}_i^+(\mathcal{N}))^2$ which is obtained by the expansion of Eq.~\eqref{E5} versus the 
 truncation orders $\mathcal{N}$ for a spin chain of lengths $L=7,9,11$ and with different localization lengths $\xi=1,2,3$  correspondingly which is averaged over different disorder configurations.}
	\label{FG2}
\end{figure}

 \begin{figure}[t!]
	\center{\includegraphics[width=1.0\linewidth]{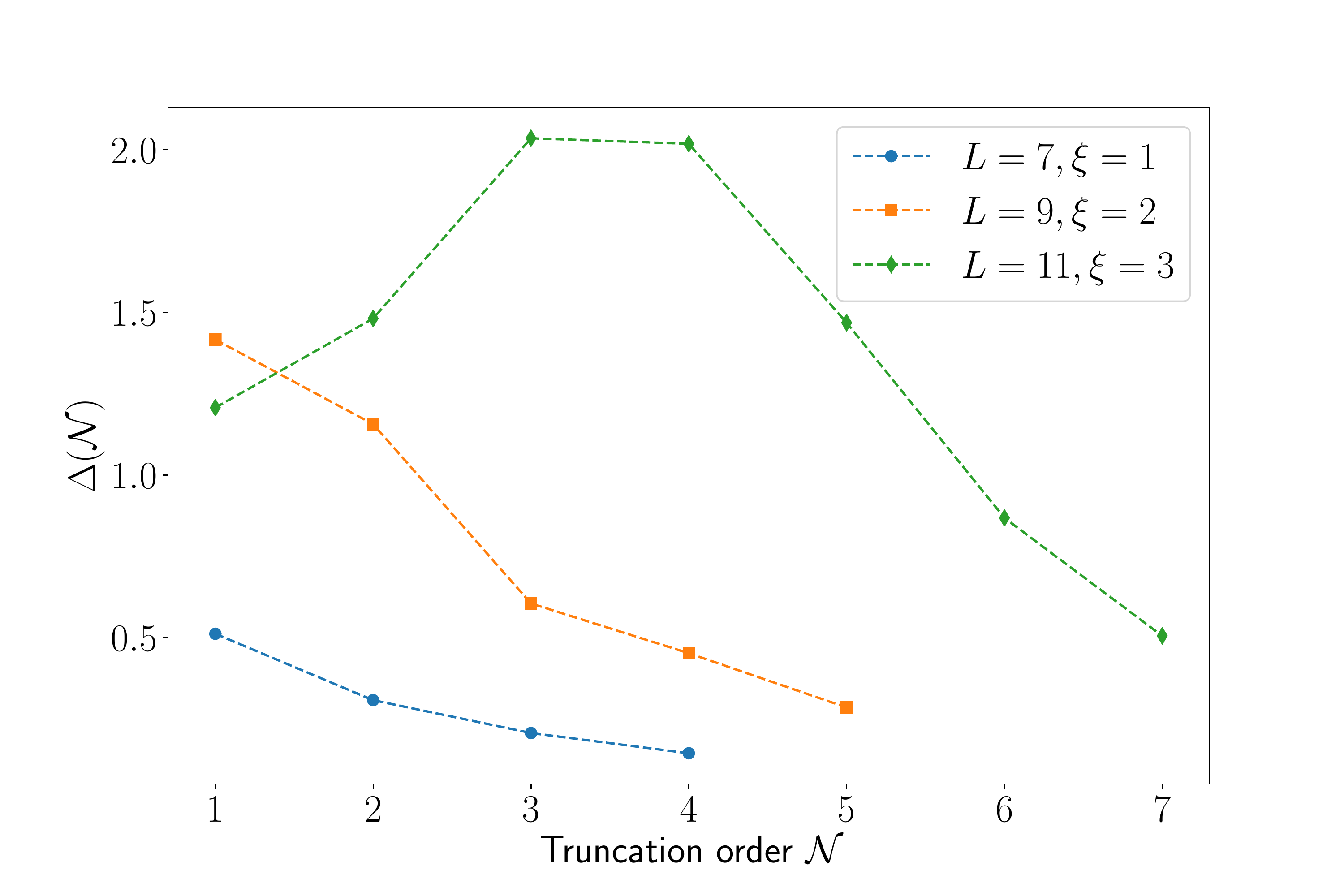}} 
	\caption{The relative Frobenius norm of the commutator of approximated-LIOM operators $\tau_i^z$ in Eq.~\eqref{E4} obtained using the expansion of Eq.~\eqref{E5} and the Hamiltonian which is defined as $\Delta (\mathcal{N})$ in Eq.~\eqref{E30} versus the 
 truncation orders $\mathcal{N}$ for a spin chain of lengths $L=7,9,11$ and with different localization lengths $\xi=1,2,3$  correspondingly  which is averaged over different disorder configurations.
}
	\label{FG3}
\end{figure}

To check the impact of the truncation order more precisely, we have also studied the behavior of $||(\tilde{\tau}_i^+)^2||^2_F$ which is shown in Fig.~\ref{FG2}. 
In-principal, $\tilde{\tau}_i^+$ should satisfy the fermionic algebra and hence, $(\tilde{\tau}_i^+)^2=0$.
However, the plots of Fig.~\ref{FG2} show that the higher-order corrections induce a large deviation from their fermionic property for ($\mathcal{N}<(\xi/a)$) which means that the dressing effects of the larger number of pairs of particle-hole excitation may significantly change the picture of quasiparticles associated with LIOMs and their perturbative expansion is not always a trivial one.

The second quantity which we are interested in  is the relative Frobenius norm of the commutator of approximated-LIOM operators and the Hamiltonian which is defined as
\be
\label{E30}
\Delta (\mathcal{N})= \frac{||[\mathcal{H},\tau_{i}^z(\mathcal{N})]||^2_F}{||\tau_{i(\text{Exact})}^z||^2_F},
\ee
where $\tau_{i}^z(\mathcal{N})$ can be obtained by plugging the result of Eq.~\eqref{E5} in Eq.~\eqref{E4} and $\tau_{i(\text{Exact})}^z$ is the exact LIOM operator obtained using the method of Ref.~\cite{Amini2022}.
As it is obvious again in Fig.~\ref{FG3} that the localization length is the most important parameter in the expansion of LIOMs. Because the contributions come from the sectors in which the number of pairs of particle-hole excitation is smaller than the localization radius plays a crucial role in the construction of LIOM operators. 
This means that writing a perturbative expansion for the construction of LIOMs (or corresponding quasiparticle operators) in the MBL phase is always a tricky business since the contributions come from different sectors strongly depend on the localization length of the system.


\section{Summary} \label{Sec.IV}
In conclusion, we studied the mechanism of dressing the quasiparticle operator associated with LIOMs with different numbers of pairs of particle-hole excitation in the MBL phase of a disordered spin chain. 
By expanding the LIOM creation operator in terms of different single-particle operators dressed by different numbers of particle-hole excitations and analytical derivation of the coefficients of the expansion, we explicitly obtained the contributions of the sectors with different excitations. We have performed an order-by-order expansion which allows identify precisely the role played by higher order corrections.  
Our analysis showed that the key parameter to determine the optimum order of truncation for this expansion is the localization length of the system. 

\begin{acknowledgments}
MA acknowledges the support of the Abdus Salam (ICTP) associateship program.
\end{acknowledgments}



\begin{thebibliography}{}

\bibitem{RMP} D. A. Abanin, E. Altman, I. Bloch and M. Serbyn, Many-body localization, thermalization, and entanglement, Rev. Mod. Phys. 91, 021001 (2019).
\bibitem{Mirlin2005} I. V. Gornyi, A. D. Mirlin, and D. G. Polyakov, Interacting electrons in disordered wires: Anderson localization and low-$T$ transport, Phys. Rev. Lett. 95, 206603 (2005).
\bibitem{BAA} D. Basko, I. Aleiner, and B. Altshuler, Metal-insulator transition in a weakly interacting many-electron system with localized single-particle states, Ann. Phys. (NY) 321, 1126 (2006).
\bibitem{Anderson} P. W. Anderson, Absence of diffusion in certain random lattices, Phys. Rev. 109, 1492 (1958).
\bibitem{Oganesyan} V. Oganesyan and D. A. Huse, Localization of interacting fermions at high temperature, Phys. Rev. B 75, 155111 (2007).
\bibitem{Luitz} D.J. Luitz, N. Laflorencie, F. Alet,  Many-body localization edge in the random-field Heisenberg chain, Phys. Rev. B 91,  081103(R) (2015).
\bibitem{Ex1} M. Schreiber, S. S. Hodgman, P. Bordia, H. P. Luschen, M. H. Fischer, R. Vosk, E. Altman, U. Schneider, and I. Bloch, Observation of many-body localization of interacting fermions in a quasirandom optical lattice, Science 349, 842–845 (2015).
\bibitem{Ex2} J.-y. Choi, S. Hild, J. Zeiher, P. Schauss, A. Rubio-Abadal, T. Yefsah, V. Khemani, D. A. Huse, I. Bloch, and C. Gross, Exploring the many-body localization transition in two dimensions, Science 352, 1547–1552 (2016).
\bibitem{Ex3} J. Smith, A. Lee, P. Richerme, B. Neyenhuis, P. W. Hess, P. Hauke, M. Heyl, D. A. Huse, and C. Monroe, Many-body localization in a quantum simulator with programmable random disorder, Nature Physics 12, 907–911 (2016).
\bibitem{Ex4} P. Roushan, C. Neill, J. Tangpanitanon, V. M. Bastidas, A. Megrant, R. Barends, Y. Chen, Z. Chen, B. Chiaro, A. Dunsworth, and et al., Spectroscopic signatures of localization with interacting photons in superconducting qubits, Science 358, 1175–1179 (2017).
\bibitem{Ex5} K. Xu, J.-J. Chen, Y. Zeng, Y.-R. Zhang, C. Song, W. Liu, Q. Guo, P. Zhang, D. Xu, H. Deng, K. Huang, H. Wang, X. Zhu, D. Zheng, and H. Fan, Emulating many-body localization with a superconducting quantum processor, Phys. Rev. Lett. 120, 050507 (2018).
\bibitem{Znidaric} M. Znidaric, T. Prosen and P. Prelovsek, Many-body localization in the Heisenberg XXZ magnet in a random field, Phys. Rev. B 77, 064426 (2008).
\bibitem{Bardarson} J. H. Bardarson, F. Pollmann and J. E. Moore, Unbounded growth of entangle- ment in models of many-body localization, Phys. Rev. Lett. 109, 017202 (2012).
\bibitem{Gopalakrishnan} S. Gopalakrishnan, M. Muller, V. Khemani, M. Knap, E. Demler, and D. A. Huse, Phys. Rev. B 92, 104202 (2015).
\bibitem{Syrbin20} M. Serbyn, Z. Papic and D. A. Abanin, Universal slow growth of entanglement in interacting strongly disordered systems, Phys. Rev. Lett. 110, 260601 (2013).
\bibitem{Serbyn13} M. Serbyn, Z. Papic, and D. A. Abanin, Local Conservation Laws and the Structure of the Many-Body Localized States, Phys. Rev. Lett. 111, 127201 (2013).
\bibitem{Huse14} D. A. Huse, R. Nandkishore, and V. Oganesyan, Phenomenology of fully many-body-localized systems, Phys. Rev. B 90, 174202 (2014).
\bibitem{Chandran15} A. Chandran, I. H. Kim, G. Vidal, and D. A. Abanin, Constructing local integrals of motion in the many-body localized phase, Phys. Rev. B 91, 085425 (2015).
\bibitem{ROS} V. Ros, M. Müller, and A. Scardicchio, Integrals of motion in the many-body localized phase, Nucl. Phys. B 891, 420 (2015).
\bibitem{Imbrie17} J. Z. Imbrie, V. Ros, and A. Scardicchio, Local integrals of motion in many‐body localized systems, Annalen der Physik 529, 1600278 (2017).
\bibitem{Geraedts} S. D. Geraedts, R. N. Bhatt, and R. Nandkishore, Emergent local integrals of motion without a complete set of localized eigenstates, Phys. Rev. B 95, 064204 (2017).
\bibitem{Singh} H Singh, B Ware, R Vasseur, S Gopalakrishnan, Local integrals of motion and the quasiperiodic many-body localization transition, Phys. Rev. B 103, L220201 (2021).
\bibitem{Rademaker} L. Rademaker and M. Ortuño, Explicit Local Integrals of Motion for the Many-Body Localized State, Phys. Rev. Lett. 116, 010404 (2016).
\bibitem{Abanin2016} T. E. O’Brien, D.A. Abanin, G. Vidal, and Z. Papic, Explicit construction of local conserved operators in disordered many-body systems, Phys. Rev. B 94, 144208 (2016).
\bibitem{Goihl} M. Goihl, M. Gluza, C. Krumnow, and J. Eisert, Construction of exact constants of motion and effective models for many-body localized systems, Phys. Rev. B 97, 134202 (2018).
\bibitem{Peng} P. Peng, Z. Li, H. Yan, K. X. Wei, and P. Cappellaro, Comparing many-body localization lengths via nonperturbative construction of local integrals of motion, Phys. Rev. B 100, 214203 (2019).
\bibitem{Amini2022} S. Adami, M. Amini, and M. Soltani, Structural properties of local integrals of motion across the many-body localization transition via a fast and efficient method for their construction, Phys. Rev. B 106, 054202  (2022). 
\bibitem{Bera} S. Bera, T. Martynec, H. Schomerus, F. Heidrich- Meisner, and J. H. Bardarson, One-particle density ma- trix characterization of many-body localization, Ann. Phys. (Berl.) 529, 1600356 (2017).
\bibitem{Chen} C. P. Chen and H. Schomerus, Fock-space geometry and strong correlations in many-body localized systems, Phys. Rev. B 104, 205411 (2021).
\bibitem{PT1} A. Pal and D.A. Huse, Many-body localization phase transition, Phys. Rev. B 82, 174411 (2010).
\bibitem{PT2} D.  J.  Luitz, N. Laflorencie, and F. Alet,  Many-body localization edge in the random-field Heisenberg chain, Phys. Rev. B 91, 081103(R) (2015).



\end{thebibliography}
\end{document}